# Engineering Photoluminescence with Mie Voids

*Yuchao Fu[1]*, Ilia Lykov[1], Sergejs Boroviks[1], Nai-Quan Zhu[2], Tianyue Li[3], Siarhei Zavatski[1], Makhlad Chahid[4], Olivier J. F. Martin[1]**

*[1] Nanophotonics and Metrology Laboratory, Swiss Federal Institute of Technology Lausanne (EPFL); 1015 Lausanne, Switzerland.*
*[2] School of Electronic Information and Electrical Engineering, Shanghai Jiao Tong University (SJTU); 200240 Shanghai, China.*
*[3] Department of Physics, The Hong Kong University of Science and Technology; 999077, Hong Kong.*
*[4] Center of MicroNanoTechnology, Swiss Federal Institute of Technology Lausanne (EPFL); 1015 Lausanne, Switzerland.*

**Corresponding author. Email: yuchao.fu@epfl.ch ; olivier.martin@epfl.ch*

## Abstract

Spontaneous emission, as a fundamental radiative process and a versatile information carrier, plays a vital role in light-emitting devices, optical information modulation and encryption, super-resolution fluorescence imaging. Engineering the photonic environment surrounding emitters enables control over their emission properties. However, simultaneously achieving precise engineering of both excitation enhancement and quantum-yield modulation at the nanoscale remains elusive, highlighting substantial room for advancing the precise orchestration of photoluminescence. Here, we introduce silicon Mie voids — air-defined cavities that invert the conventional solid-particle geometry — to achieve independent tuning of photoluminescence within a single subwavelength unit, while minimizing optical losses. Full-wave simulations and experiments on both gradient and uniform Mie-void arrays jointly validate this quantitative framework for spontaneous emission tuning, which disentangles excitation enhancement arising from local field confinement in air and quantum-yield enhancement resulting from strengthened emitter-resonator coupling, while confirming the accelerated radiative decay enabled by the modified optical LDOS. Leveraging this flexible mechanism, we realize a multimodal nanophotonic pattern with near–diffraction-limited pixels that encode the EPFL logo in the bright field and the SJTU logo in both dark field and photoluminescence maps. These results establish Mie voids as a powerful platform for high-density multiplexed encrypted displays and open new avenues for advancing state-of-the-art nanophotonic devices.

## Introduction

Tailoring light–matter interactions at the nanoscale is a central aspect of modern photonics, enabling unprecedented advances in spontaneous emission control, chiral luminescence, and nonlinear optics[1-3]. In the pursuit for miniaturization, high-index dielectric nanostructures offer a highly versatile alternative to plasmonic systems. Nanostructured high-index materials, such as silicon (Si) and gallium arsenide (GaAs) and other semiconductors, enable strong confinement of optical modes with reduced absorption losses, generating high-purity structural colors and exhibiting novel optical states such as anapoles, Fano resonances, and bound states in the continuum (BICs)[4-10]. These effects originate from the interplay between electric and magnetic multipoles induced by optically driven displacement currents inside high-index dielectric particles[5,7]. Furthermore, they enable scattering anisotropy, exemplified by forward–backward asymmetry under Kerker conditions[11-13]. The optical response of all-dielectric nanostructures can be well described in terms of the Mie theory[14]. A particularly compelling recent advancement has been the emergence of all-dielectric *Mie voids* — nanoscale air-filled cavities embedded in high-index dielectrics — which invert the traditional solid-particle geometry while preserving multipolar Mie resonances[15]. By displacing the resonant field maxima from optically lossy material regions into low-index air, all-dielectric Mie voids reduce absorptive losses to a negligible level, while simultaneously enabling strong field confinement in air and extending the resonant optical responses toward shorter wavelengths, including the blue and even ultraviolet regimes[15,16]. Compared to traditional metasurfaces composed of solid Mie scatterers[4,5,17], Mie voids provide an air-defined nanophotonic platform that introduces new degrees of freedom for engineering light-matter interactions at the nanoscale. Although this field of research is in its infancy, the potential of Mie voids for practical applications seems tremendous and has recently been demonstrated in nanoplastics detection and polarization-sensitive photonic device design[18,19]. Here, we report for the first time that Mie voids can also modify the local density of optical states (LDOS) and control photoluminescence in emitter-loaded photonic platforms.

Spontaneous emission represents an intrinsic property of isolated emitters[20,21], and it can be profoundly modified by virtue of the Purcell effect, i.e. by tailoring their surrounding photonic environment[17,22]. According to Purcell's theory, the spontaneous emission rate is proportional to the LDOS, and thus can be enhanced by embedding the emitter in a resonant structure with high quality factor ($Q$) and small mode volume ($V_m$)[23-26]. Plasmonic nanostructures initially led this field by providing deeply subwavelength field confinement and large LDOS, thereby yielding large Purcell factors; however, their observable quantum efficiencies are often compromised by quenching induced by intrinsic ohmic losses[17,27-30]. On the other hand, photonic crystal slabs enable record-high fluorescence enhancement (nearly 3000 -fold) through combined resonant excitation and directional extraction[31]. Yet, both approaches demand sophisticated nanofabrication and offer limited flexibility for spatially resolved emission control[31-33]. In contrast to their plasmonic counterparts, all-dielectric Mie resonators are comparatively facile to fabricate thanks to their larger feature sizes and can simultaneously support directional emission, flexible spatial arrangement, and lifetime shortening with reduced contributions from non-radiative decay[12,34,35]. Recent progress in all-dielectric nanophotonic platforms, such as Fano metasurfaces and quasi-BIC metasurfaces, has enabled nonlocal ultrahigh-Q resonances on large-area periodic structures that deliver strong Purcell enhancement together with angular control over far-field emission[30,31,36-38]. Altogether, these studies pave the way for chip-scale and room-temperature emitters with engineered radiative lifetimes, directional luminescence, and on-demand spectral shaping, extending spontaneous emission control to practical applications in nanolasers and super-resolution imaging[39-43].

Despite recent advances in Mie resonators and all-dielectric metasurface designs, these platforms still face some fundamental challenges, including the difficulty of simultaneously achieving precise and flexible control over both excitation enhancement and quantum yield enhancement, as well as spatially resolved emission control at the single-resonator scale. A particularly important limitation of most solid-particle Mie resonators is that their effective mode volume is largely confined within the high-index dielectric material[2,5,44], leading to intrinsically weak resonant coupling with emitters residing in the surrounding low-index environment. In principle, the emitters can be embedded inside the particles[45], however this involves complex chemistry protocols and limits the potential applications. Here, we propose the use of Mie voids to overcome these constraints through their complementary geometry, which localizes resonant electromagnetic fields predominantly in air and thereby facilitates stronger coupling of an external emitter to cavity modes while providing open radiation channels. Although Mie voids have been explored for enhancing light–matter interactions with two-dimensional materials[46], their potential for photoluminescence control with subwavelength spatial resolution remains largely unexplored. Fluorophore-loaded Mie-void hybrid photonic systems therefore unlock the potential for precisely orchestrating photoluminescence responses. To the best of our knowledge, there is no systematic study of how Mie voids tune photoluminescence compared to solid dielectric resonators, and a quantitative framework linking the geometric flexibility of Mie voids, local-field-driven excitation enhancement, and resonance-mediated quantum-yield enhancement is still lacking.

In this work, we focus on establishing a flexible and precise platform for tailoring photoluminescence within the smallest possible submicron void volume. We compare the emitter-accessible mode volume, hotspot distributions, and Purcell factors of conventional solid Mie particles versus air-defined Mie voids[15,47,48], revealing that the latter offer greater emitter accessibility and enhanced optical LDOS. We also analyze the photoluminescence tuning effect of Mie voids on emitters from two complementary perspectives[49]: excitation enhancement, driven by local field concentration within the void, and quantum yield enhancement, quantified via the Purcell factor, dielectric absorption, and far-field radiation efficiency, which together determine the observable photoluminescence modulation. To experimentally validate this framework, we fabricated both gradient and uniform Mie void arrays using focused ion beam (FIB) milling into amorphous silicon substrates. We demonstrate that silicon-based dielectric Mie voids have the unique physical capability of enabling flexible photoluminescence modulation within a single submicron unit without inter-void crosstalk, with gradient-array-induced emission showing close agreement with theoretical predictions. In uniform Mie void arrays, we observe accelerated radiative decay rates enabled by Mie-resonant coupling, and quantitatively validate a strong linear correlation between the model-predicted photoluminescence tuning factor and the experimentally measured emission intensity. Building upon this analytical framework established in silicon Mie voids, we further design and realize a multimodal nanophotonic structure with single-void-per-pixel spatial resolution that displays distinct patterns under white-light microscopy (bright- and dark-field imaging) and photoluminescence maps. This work establishes a theoretical foundation for the effective tunability of photoluminescence, experimentally validates the versatility and spatial programmability of all-dielectric Mie voids, and opens a new and adaptable platform for applications in light-emitting devices, encrypted display technologies [17,37,48,50-52].

## Results

### Mie resonators

We begin by systematically comparing Mie resonators to explore the potential of Mie voids, considering the elementary cases of two prototypical configurations in vacuum: an individual Mie nanoantenna and a single Mie void. Silicon, owing to its high refractive index, low cost, and compatibility with scalable industrial nanofabrication techniques, stands out as a prime candidate for realizing Mie-type nanophotonics[53-56]. The dielectric medium is assumed to be homogeneous and isotropic, with its optical constants (refractive index and extinction coefficient) provided in Supplementary Section S1. The conventional Mie nanoantenna is modeled as a cylindrical silicon particle with a dipole emitter positioned adjacent to its surface (Figure 1A). By contrast, the Mie void is represented by a cylindrical air cavity etched into silicon (Figure 1B), a geometry that is both experimentally feasible and optically accessible in the visible wavelength range, with a dipole emitter embedded within the void. The influence of the Mie resonator on the emitter can be analyzed from the confinement of the local electromagnetic field at the excitation wavelength and the resonant coupling between the emitter and the Mie resonator at the emission wavelength.[49] Both effects manifest respectively as excitation enhancement and quantum yield enhancement.

At the excitation wavelength, the near-field distributions and mode volumes are computed and compared under $y$-polarized plane-wave illumination for both Mie resonators. As shown in Figure 1C, for two orthogonal cross-sections, the incident electric field is predominantly confined within the high-index dielectric of the Mie particle or strongly pinned near its surface[10]. This confinement indicates that the local electric field rapidly decreases to half of its maximum value as the emitter moves just $\sim$40 nm away from the nanoantenna[49]. This leads to an extremely small emitter-accessible mode volume (Figure 1A, upper-right inset; Eq. (S.28)), thereby limiting the spatial extent of emitter–resonator interaction. In contrast, the near-field distribution of the cylindrical silicon Mie void is shown in Figure 1D. The Mie void resonator concentrates the incident electric field within the low-index air cavity and extends into the surrounding region above the interface, which is fully accessible for emitters and minimizes energy dissipation inside the dielectric. The mode volume of the Mie void is increased by approximately two orders of magnitude compared to a solid nanoantenna with the same geometry (Figure 1B, upper-right inset). Since the local excitation field hotspots are less tightly confined than in solid Mie particles, the Mie void provides a more uniform field distribution and extends the accessible interaction region farther away from the air-dielectric interface. Furthermore, as shown in Figure 1E, the Purcell-factor map demonstrates that in solid Mie particles, resonant coupling with external emitters placed more than $\sim$20 nm away decreases rapidly, thereby limiting the extent of emission control achievable in solid Mie particles[49]. In contrast, the emission-related multipolar coupling of the Mie void is strongly localized within the air cavity and extends above the interface[4,15]. As shown in Figure 1F, pronounced regions with increased Purcell factor emerge near the center of the Mie void, enabling efficient resonant coupling for emitters located within the cavity and away from the silicon–air interface, while the presence of open radiation channels facilitates control over the apparent quantum yield observed in the far field.

Emitters located inside or directly adjacent to the dielectric are typically subject to non-radiative quenching (especially, if they are polarized orthogonally to the interface) and contribute weakly to far-field emission[57]. The insets at the bottom of Figure 1A and Figure 1B further illustrate the variation of excitation field and Purcell factor along the $y$-axis direction, where the shaded green region denotes the emitter-accessible air region outside the dielectric. Compared to the rapidly decaying field and even more rapidly decreasing Purcell factor away from the surface of a Mie particle, the Mie void supports both excitation and coupling hotspots near the cavity center, with minimal field penetration into the dielectric. Therefore, Mie voids provide significantly improved emitter accessibility, enabling more spatially uniform and effective emission control. Importantly, since the mode supported by the void is highly sensitive to its geometry, this configuration also offers an effective mechanism for spatially resolved luminescence tuning at the single-resonator level, independent of neighboring nanostructures. By tailoring the geometry of individual Mie voids, one can modulate the distinct resonant cavity modes confined within their respective air regions, thereby enabling spatially independent control of emitters residing inside each void. This principle lays the physical foundation for high-spatial-resolution photoluminescence encoding patterns, free from crosstalk between adjacent pixelated voids, as will be illustrated in the final section of this work.

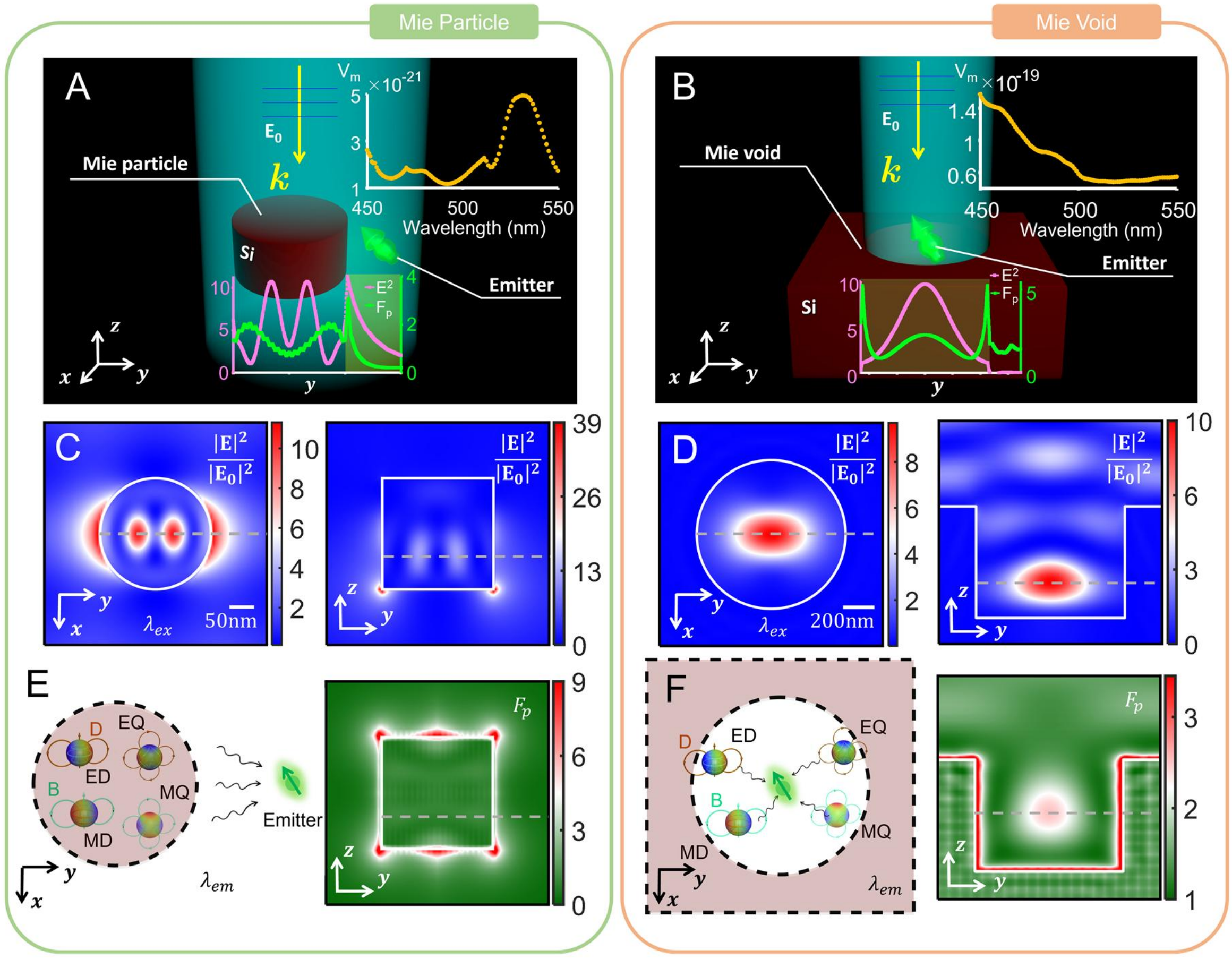

**Fig. 1. Silicon-based Mie resonators reveal local field confinement and resonant coupling with the emitter. (A)** Schematic of a dipole emitter placed adjacent to a cylindrical silicon Mie particle (radius = 100 nm, height = 200 nm) and illuminated by a plane wave. **(B)** Schematic of an emitter embedded in a cylindrical Mie void (radius = 460 nm, height = 430 nm) etched into silicon under the same illumination. The insets in (A) and (B) show the mode volume analysis and emitter accessibility for the two Mie resonators. **(C)** Near-field intensity distributions $|\mathbf{E}|^2/|\mathbf{E}_0|^2$ in the $xy$- and $yz$-planes for a cylindrical silicon Mie particle, and **(D)** for a cylindrical silicon Mie void, respectively. The excitation wavelength is $\lambda_{ex} = 488$ nm. The Mie void exhibits a mode volume approximately two orders of magnitude larger and supports localized field hotspots away from the dielectric–air interface. **(E)** and **(F)** show cross-sectional maps of the Purcell factor, illustrating the dependence of resonant coupling strength on the spatial position of the emitter for the Mie nanoantenna and Mie void, respectively. The emission wavelength is $\lambda_{em} = 520$ nm. The reported Purcell factors are averaged over three mutually orthogonal dipole orientations ($x$, $y$, and $z$). The Mie void supports pronounced resonant coupling hotspots near the center of the cavity. Note that in (E) and (F), the electric dipole (ED), magnetic dipole (MD), electric quadrupole (EQ), and magnetic quadrupole (MQ)

contributions are shown separately for clarity, although these modes coexist and spatially overlap in practice.

**Photoluminescence tuning**

To quantitatively investigate the photoluminescence tuning capabilities of an air-defined nanophotonic environment, we consider a Mie void resonator etched into a high-index silicon substrate (Figure 2A). Under realistic fabrication conditions, such Mie voids typically exhibit a truncated conical profile characterized by a height $h$ and upper/lower radii $R_1$ and $R_2$. For analytical clarity and computational efficiency, we start by modeling the Mie void as a vertically symmetric cylindrical cavity with an emitter placed at its geometric center. This configuration supports strong Mie resonances that confine the electric field predominantly within the low-index air region and facilitate efficient resonant coupling with embedded emitters. When illuminated by a normally incident plane wave polarized along the $y$-direction and propagating along the $z$-direction, the emitter undergoes optical excitation governed by the local electric field $\mathbf{E_d}(\boldsymbol{r_0}, \omega)$ at the emitter location. A detailed derivation of the excitation enhancement factor $F_{ex}$ based on this field confinement is provided in Supplementary Section S2. In practice, emitters are typically distributed randomly and uniformly throughout the air-filled volume of the Mie void. To evaluate the overall excitation enhancement within the Mie void, we adopt a spatially averaged formulation of the enhancement factor given in Eq. (S.14),

$$\bar{F}_{ex}(\boldsymbol{r_0}, \omega) = \frac{1}{V}\int_V \frac{|\mathbf{E_d}(\boldsymbol{r_0}, \omega)|^2}{|\mathbf{E_0}(\boldsymbol{r_0}, \omega)|^2} dV\,, \tag{1}$$

where the integration volume $V$ refers to the region encompassing the Mie void and an overlying layer of height $t$, which corresponds to the thickness of the emitter-containing medium. The complex amplitude of the incident electric field at the location $\boldsymbol{r_0}$ of the emitter is $\mathbf{E_0}(\boldsymbol{r_0}, \omega)$ and $\omega$ is the angular frequency. Building on the above framework, we numerically computed the near-field distributions for Mie voids across a broad range of geometric parameters, and subsequently evaluated their spatially averaged excitation enhancement factors $\bar{F}_{ex}$ (Eq. 1), as shown in Figure 2B. The small-radii Mie voids (below ~400 nm) exhibit a strong dependence of $\bar{F}_{ex}$ on the cavity depth. Values of $\bar{F}_{ex} < 1$ can even be observed for deep and narrow cavities, indicating that such void geometries fail to achieve effective field confinement therein. In such cases, the cavity geometry restricts the excitation channels and thereby suppresses energy delivery to the emitter. By contrast, in large-radii Mie voids, the excitation enhancement factor becomes relatively insensitive to the void depth and approaches a plateau value of ~1.5. This behavior suggests that lateral confinement dominates the field enhancement mechanism in large-diameter cavities, and that the excitation enhancement becomes less sensitive to the vertical geometry once stable resonance conditions are satisfied, implying a geometric saturation of the local field confinement effect.

The preceding analysis focused solely on the excitation enhancement effect induced by the Mie void under optical illumination. However, a comprehensive understanding of photoluminescence tuning also requires evaluation of the resonator-mediated enhancement of spontaneous emission, namely, the Purcell effect. To isolate this contribution, we eliminate the external illumination and consider an emitter undergoing spontaneous decay into three channels (Figure 2A): intrinsic nonradiative decay ($\gamma_{nr,0}$), radiative emission to the far field ($\gamma_{rad}$), and additional absorption in the surrounding lossy dielectric ($\gamma_{abs}$). Supplementary Section S3 provides a quantitative framework of the resonant coupling and emission dynamics using three dimensionless factors: the Purcell factor ($F_p$), the far-field radiation factor ($\mu$), and the dielectric absorption factor ($\mu_1$). The energy-based definitions and integration methods for extracting these parameters from simulations are also detailed therein. The Purcell factor describes the modification of the emitter decay rate induced by changes in the optical LDOS, the dielectric absorption factor quantifies the portion of energy dissipated within the surrounding lossy dielectric medium, and the far-field radiation factor denotes the fraction of energy radiated to the far field and collected by a far-field receiver. The combined effect of these three terms governs the resonance-modified quantum yield enhancement factor (Eq. (S.29)),

$$F_q(\boldsymbol{r_0}, \omega) = \frac{q_d}{q_0} = \frac{F_p - \mu_1}{q_0(F_p - 1) + 1} = \frac{\mu}{q_0(F_p - 1) + 1}, \tag{2}$$

where $q_0$ and $q_d$ are the intrinsic and resonance-modified quantum yields, respectively. This framework enables us to systematically evaluate the geometric modulation of the emitter-resonator interactions. Figure 2C presents calculations of a dipole emitter placed at a distance of 150 nm from the center of a 100 nm-diameter silicon Mie particle. We computed the Purcell factor, far-field radiation factor, and dielectric absorption factor for two orthogonal dipole orientations: one aligned vertically (left in Figure 2C) and the other horizontally across the particle center (right in Figure 2C). In both cases, the far-field receiver surface A is defined as a spherical shell enclosing the resonant system. The results confirm the energy-conservation relation $F_p = \mu + \mu_1$, thereby validating the reliability of the simulations and allowing for simplification in subsequent analyses. Specifically, this relation eliminates the need for full-volume absorption integration ($P_{abs}$) to determine $\mu_1$ in Eq. 2, reducing the procedure to the computation of the surface-integrated radiated power ($P_{rad}$) for evaluating $\mu$. Moreover, Figure 2C reveals a strong dependence of the quantum yield enhancement on emitter orientation, highlighting the anisotropic nature of the emitter-resonator coupling. This quantum yield enhancement equation also shows how resonant coupling in Mie resonators impacts emitters with varying intrinsic quantum yields $q_0$. Figure 2D shows the relationship between $F_q$ and $q_0$ for different combinations of Purcell and dielectric absorption factors. When $F_p > 1$, indicating intrinsic resonant enhancement, the spontaneous emission improvement becomes more pronounced for emitters with lower intrinsic quantum yields [58]. In contrast, when $F_p < 1$, $F_q$ remains consistently below unity, signifying actual emission suppression — an effect that is again more severe for emitters with lower $q_0$. To achieve substantial quantum yield enhancement ($F_q > 1$) through resonant coupling in dielectric Mie resonators, the dielectric absorption must be sufficiently low such that $\mu_1 < (F_p - 1)(1 - q_0)$. Additional calculation results with strong Purcell effect and coupling interaction between a dipole emitter and the nanophotonic resonator are provided in Supplementary Section S4.

In practical experimental setups, far-field detectors are typically positioned above the Mie-void structures. To reflect this configuration in the calculations, we define the far-field receiver surface A as an infinite horizontal plane located above the void. This setup is conceptually analogous to projecting the spherical receiver surface of a Mie particle onto a planar receiving surface (inset of Figure 2B). We conducted a systematic sweep of the geometric parameters of silicon Mie voids and computed the corresponding values of the three dimensionless factors, as summarized in Figures 2E, 2F, and 2G, respectively. Additional results with extended height variations can be found in Supplementary Section S5. These results reveal that the resonant coupling between the embedded emitter and Mie voids can be continuously tuned by varying the void radius and depth, enabling flexible control of the Purcell factor over a range from approximately 0.5 to 3.5. Since the far-field receiver plane A only accounts for upward radiation channels, the calculated apparent quantum yield enhancement factor is always less than unity (Figure 2G). Nonetheless, this analysis confirms that the resonant coupling remains the dominant mechanism governing the spontaneous emission behavior of emitters embedded within Mie voids.

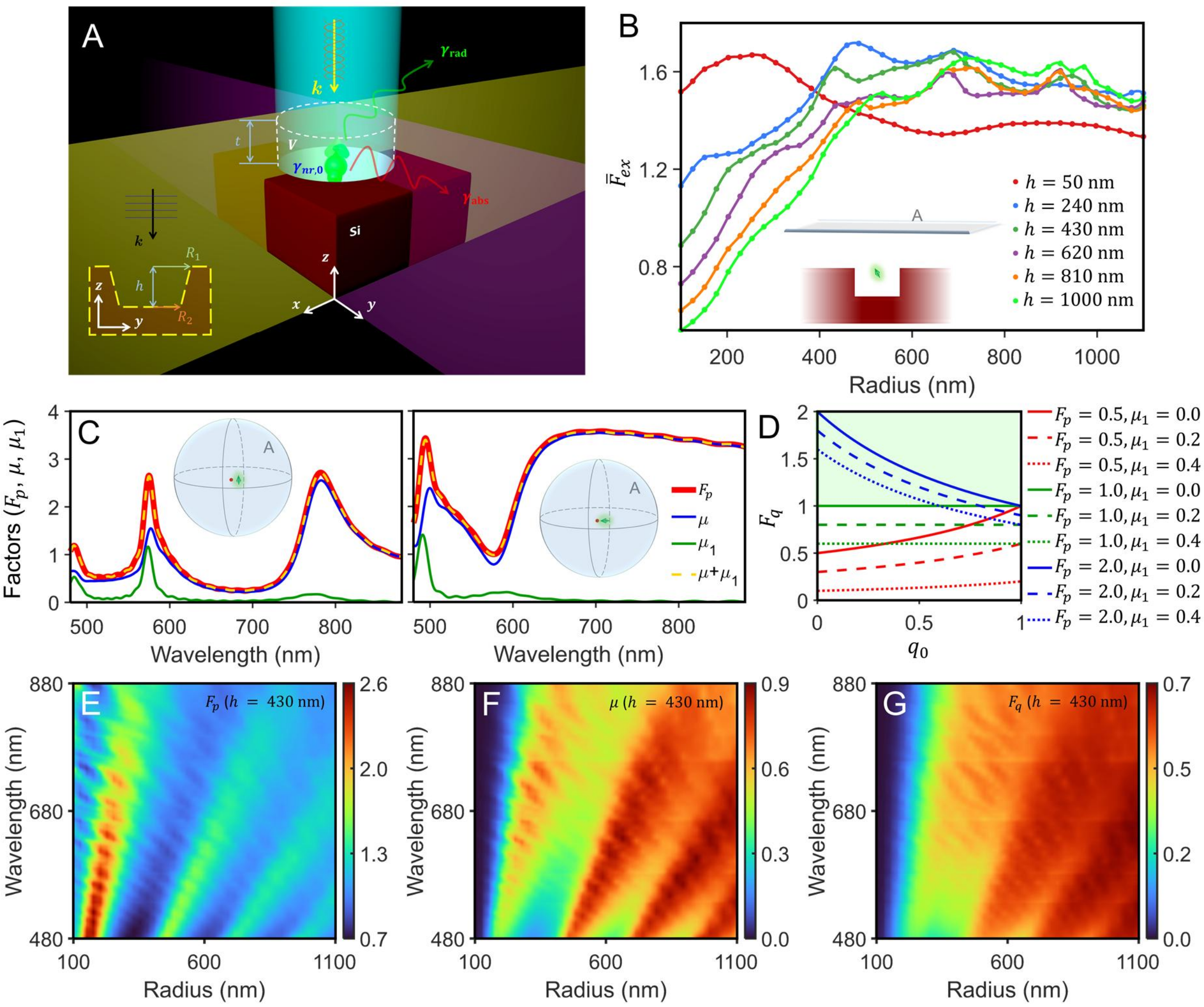

**Fig. 2. Tuning photoluminescence with all-dielectric Mie void resonators. (A)** Schematic illustration of photoluminescence processes in a silicon Mie void resonator. An emitter (e.g., a fluorophore or quantum dot) is embedded inside a Mie void. Under optical excitation from a normally incident $y$-polarized plane wave (blue beam), the emitter undergoes near-field excitation and subsequent spontaneous emission. Three decay rates ($\gamma_{rad}$, $\gamma_{nr,0}$, $\gamma_{abs}$) determine the resonator-modified quantum yield. The inset shows a truncated conical geometry more representative of realistic fabricated Mie voids, characterized by height $h$ and top/bottom radii $R_1$ and $R_2$. **(B)** Spatially averaged excitation enhancement factor $\bar{F}_{ex}$ as a function of the geometric parameters of cylindrical Mie voids. The inset illustrates the definition of the far-field receiver plane A positioned above the Mie void, used for far-field radiative energy calculations. **(C)** Energy-conservation relation for a dipole emitter placed 150 nm from the center of a 100 nm-diameter silicon Mie particle. Left: vertical dipole orientation; Right:

horizontal dipole orientation. Shown are the Purcell factor ($F_p$), far-field radiation factor ($\mu$), dielectric absorption factor ($\mu_1$), and their sum, validating the relation $F_p = \mu + \mu_1$. Insets indicate the spherical receiver surface A enclosing the particle–emitter system. **(D)** Dependence of the resonance-modified quantum yield enhancement factor $F_q$ on the intrinsic quantum yield for selected values of $F_p$ and $\mu_1$, as described by Eq. 2. The enhancement is strongly dependent on $q_0$, with low-$q_0$ emitters being more strongly influenced by resonant coupling. **(E)** Parameter sweeps of the Purcell factor, **(F)** the far-field radiation factor, and **(G)** the quantum yield enhancement factor as functions of void radius and emission wavelength, with void depth fixed at 430 nm. The far-field collection is restricted to upward emission.

## Experimental Results

Building on the above analytical framework, we experimentally validate the concept of engineering photoluminescence with Mie voids. We fabricated a gradient array of Mie-void resonators in amorphous silicon using FIB milling to experimentally validate the capability of Mie voids to modulate the fluorescent molecules. The gradient array refers to a two-dimensional layout in which the milling radius of the Mie voids increases linearly along the horizontal (row) direction, while the milling depth increases linearly along the vertical (column) direction. Figure 3A presents scanning electron microscope (SEM) images of the fabricated sample: the central panel shows a top-down view of the full array, while the four corner insets display oblique-angle SEM images taken at a 32° tilt, providing clear visualization of the gradual variations in both radius and depth of the cylindrical Mie voids across orthogonal directions. To precisely characterize the geometrical profile of the fabricated voids, we further performed atomic force microscopy (AFM) characterization of the array sample (Supplementary Section S8). In fact, the fabricated voids exhibit truncated conical geometries; for consistency with numerical modeling, their effective radius is defined as the average of the top and bottom radii. The measured radii vary from 160 nm (leftmost column) to 360 nm (rightmost column), and the depth increases from 30 nm (bottom row) to 770 nm (top row). Figures 3B and 3C show the bright-field and dark-field optical microscope images of the pristine Mie-void gradient array, respectively. While similar reflectance results have been reported previously[15], we reproduce them here to establish a comprehensive foundation for studying encoded nanophotonic patterns, and to adopt bright-field reflection imaging as one of the display modalities in the following sections. These images reveal that air-defined Mie voids support geometry-dependent reflected and scattered colors, with spatially varying reflection and scattering spectra clearly discernible across the array. Notably, even two Mie-void pixels that exhibit markedly different spectral features in the bright-field image can appear similarly colored under dark-field observation conditions. This phenomenon offers additional flexibility in designing subwavelength-encoded patterns with high spatial resolution, enabling multiple distinct spectral responses to be encoded within a single nanophotonic pixel.

In subsequent experiments, a 10% polyvinyl alcohol (PVA) solution is employed as the host matrix for the emitters and spin-coated onto the surface of the Mie-void gradient array, forming a conformal PVA-coating layer of uniform thickness (Figure 3D). The thickness and refractive index of this PVA film were characterized using spectroscopic ellipsometry, with detailed measurements provided in Supplementary Section S1. Since the refractive index of 10% PVA is also significantly lower than that of amorphous silicon[59], the coated voids can effectively retain optical characteristics similar to dielectric–air Mie resonators, however, as will be shown later, their resonant spectral positions experience noticeable shifts due to the change in refractive index within the void. In this experiment, we chose fluorescein isothiocyanate (FITC) molecules as representative fluorophores. These fluorophores are uniformly dispersed and randomly oriented throughout both the Mie voids and the overlying PVA-coating layer. For simulation, this distribution is modeled as a dipole cloud comprising an ensemble of randomly oriented dipole emitters distributed throughout the molecule-filled volume (Figure 3E, Supplementary Section S6). The collective emission from this dipole cloud within a single Mie void can be characterized by the averaged resonance-modified quantum yield enhancement factor $\bar{F}_q$, expressed in Eq. 3. This averaging procedure involves not only spatial integration over the emitter volume, but also spectral wavelength integration across the emission bandwidth and angular integration over all possible emitter orientations in $4\pi$ steradians,

$$\bar{F}_q = \frac{1}{4\pi V \Delta\lambda} \int_V dV \int_{\lambda_0}^{\lambda_0+\Delta\lambda} d\lambda \int_0^{2\pi} d\phi \int_0^{\pi} \frac{q_d}{q_0} \sin\theta \, d\theta \, , \quad (3)$$

where the integration limit $\lambda_0$ to $\lambda_0 + \Delta\lambda$ corresponds to the full width at half maximum (FWHM) of the fluorophore spectrum, which for FITC emission spans 510 nm– 540 nm. The angles $\phi$ and $\theta$ represent the two orthogonal angular components (the azimuthal and polar angles) defining the dipole orientation in three-dimensional space. Figures 3F and 3G show the bright-field and dark-field optical microscope images, respectively, of the Mie void gradient array after PVA coating. Compared to their counterparts in air, the resonant color of each void indicates a systematic shift of the resonant wavelength.

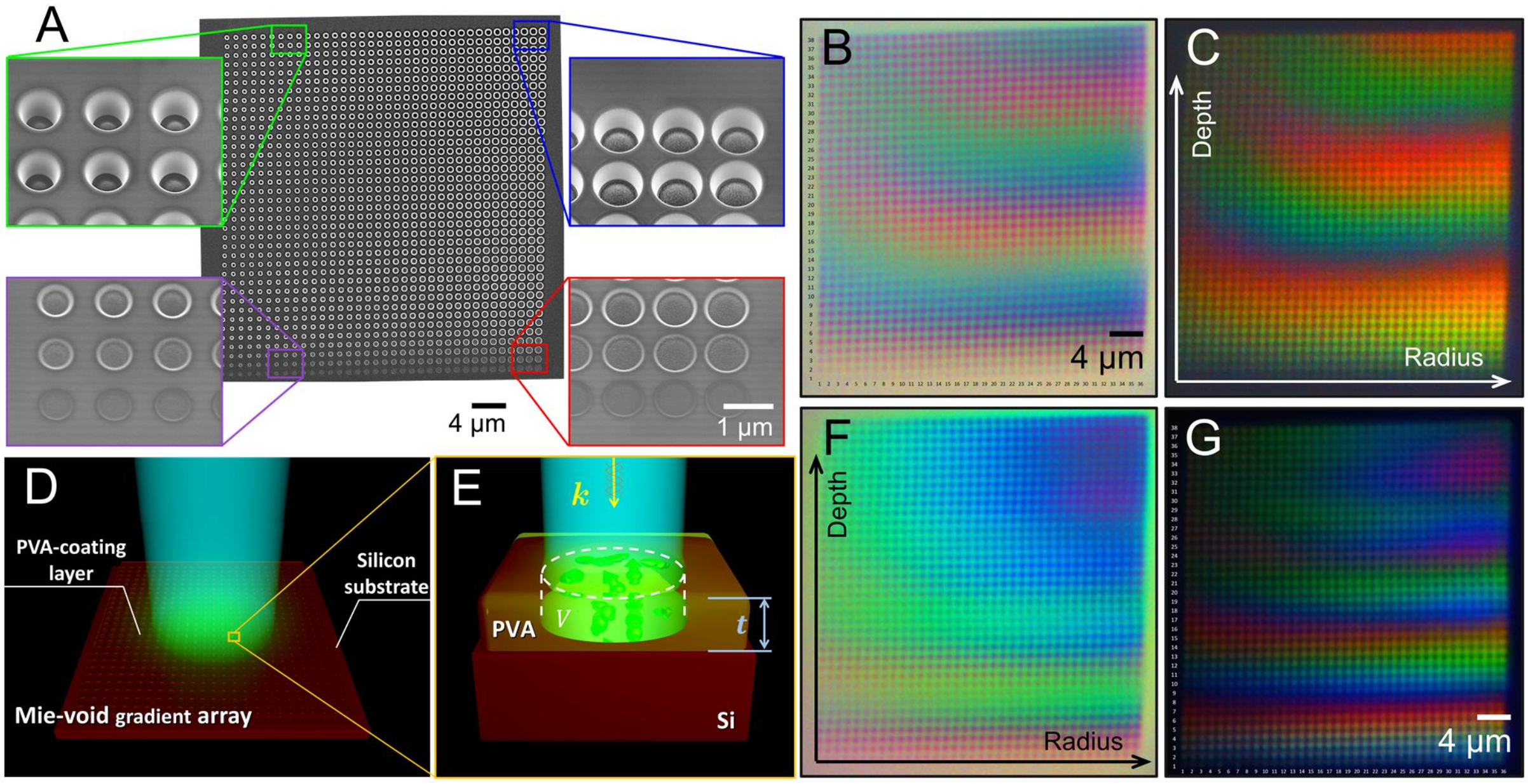

**Fig. 3. Gradient array of silicon Mie voids without and with PVA coating. (A)** SEM images of the fabricated Mie-void gradient array. The central panel shows a top-down view of the entire array, with four corner insets displaying 32°-tilted SEM views of representative voids to reveal their depth profiles. The array is patterned on a square lattice with a periodicity of 1200 nm. The nanovoid depth varies from 30 nm to 770 nm across 38 rows, and the radius varies from 160 nm to 360 nm across 36 columns. **(B)** Bright-field and **(C)** dark-field optical microscope images of the gradient array. The Mie voids exhibit distinct reflectance and scattering spectral resonances that vary across the array, as indicated by the color variations observed for both imaging modes. **(D)** Schematic of the array sample spin-coated with a transparent PVA (10% w/v) thin film uniformly embedding fluorophores and **(E)** magnified view of a single Mie void within the PVA-coating layer, showing a volumetric distribution of randomly oriented emitters extending from the void base to the film surface. **(F)** Bright-field and **(G)** dark-field optical microscope images after applying the PVA coating. The introduction of the low-index PVA layer causes a measurable spectral shift in the resonance of the Mie voids due to the modified local photonic environment in both imaging modes.

Next, we characterized the photoluminescence tuning effect in the gradient array sample using spatially resolved spectroscopy and time-correlated single-photon counting (TCSPC). The schematic of the optical setup is illustrated in Figure 4A (with detailed components and configuration provided in Supplementary Section S7). A 488 nm pulsed laser was employed to excite the fluorophores, and the emitted fluorescence was collected using both a spectrograph and a single-photon counting detector, enabling measurements of emission spectra, fluorescence lifetimes, and spatially resolved photoluminescence maps (see Materials and Methods for details). We numerically evaluated the spatially averaged excitation enhancement factor $\bar{F}_{ex}$ of PVA-coated Mie voids with varying radius and depth according to Eq. 1, using a 10% PVA coating layer with a thickness of 565 nm, identical to that in the experiments. For reference, we also calculated $\bar{F}_{ex,b}$, the corresponding excitation enhancement factor in the PVA layer on a bare silicon substrate. Figure 4B presents the resulting relative excitation enhancement $\bar{F}_{ex}/\bar{F}_{ex,b}$ for each combination of Mie void dimensions, comparing the enhancement achieved within the structured void to that on a bare silicon surface coated with the same PVA layer. In an analogous manner, we computed the resonance-modified quantum yield enhancement factor $\bar{F}_q$ for emitters embedded in the PVA-coated Mie voids of different sizes using Eq. 3, and compared it to the corresponding value $\bar{F}_{q,b}$ for emitters uniformly distributed in the PVA layer on the bare substrate. The resulting relative quantum yield enhancement $\bar{F}_q/\bar{F}_{q,b}$ is presented in Figure 4C. These results highlight the two distinct and independent mechanisms by which a Mie void tunes photoluminescence: enhancement of the local excitation field and modification of spontaneous emission via resonant coupling. Both effects are essential contributors to the overall tuning effect of photoluminescence and the combined photoluminescence tuning factor $F_{pl}$ can be expressed as[49]:

$$F_{pl}(\boldsymbol{r_0}, \omega) = F_{ex}(\boldsymbol{r_0}, \omega) \cdot F_q(\boldsymbol{r_0}, \omega). \quad (4)$$

By taking the ratio between the Mie void configuration and the unstructured bare substrate, the spatially averaged relative form $\bar{F}_{pl}/\bar{F}_{pl,b}$ is obtained, as visualized in Figure 4D. This theoretical prediction is directly compared to the experimentally acquired photoluminescence map shown in Figure 4E. The white dashed box in Figure 4D outlines the experimentally accessed range of void geometries (radius and depth, Supplementary Section S8). The comparison reveals strong agreement in the spatial distribution of luminescence intensity, confirming the efficacy of the combined excitation and emission enhancement framework. Notably, compared to emitters situated on the bare silicon surface, the Mie-void configuration yields a maximum emission enhancement ratio of 2.6 and a minimum suppression ratio of 0.5, whereas the experimentally observed range of variation is narrower (approximately 0.9 – 2.5). These discrepancies are primarily attributed to fabrication imperfections and measurement uncertainties. In particular, the FIB process produces truncated conical void profiles rather than ideal cylindrical geometries (as illustrated in the inset of Figure 2A and Supplementary Figure S12), with deviations becoming especially pronounced for deep and narrow voids. Additional fabrication-induced factors, such as surface roughness, edge rounding, and contamination with gallium, further affect the optical confinement and emitter-resonator coupling. On the measurement side, ensuring conformal thickness of the PVA coating and maintaining consistency between the excitation laser spot size and the Mie void aperture are critical for minimizing deviations between experimental results and theoretical predictions.

Time-resolved TCSPC measurements were performed to evaluate the spontaneous emission lifetime reduction induced by the modified LDOS for fluorophores within the Mie voids. For comparison, a thick fluorophore-loaded PVA layer was also spin-coated onto a quartz substrate. As shown in Supplementary Figure S13, a depth-resolved photoluminescence scan confirms that the thickness of this PVA layer exceeds $10\ \mu m$. Such a thick PVA layer can be approximated as a homogeneous photonic environment with a constant refractive index, in which the optical LDOS is not modified by any resonance. During the measurements, we used a $50\times$ objective lens ($NA = 0.9$) to tightly focus the excitation laser spot onto an individual Mie void, thereby minimizing contributions from regions outside the void to the measured emission lifetime. The decay dynamics of emitters in both samples were analyzed using a single-exponential decay model over a short time scale. As shown in Figure 4F, the emission lifetime of fluorophores in the thick PVA layer is $\tau_{PVA} = 3.49$ ns (fit shown with blue line), whereas the lifetime extracted from the TCSPC data for fluorophores within a single Mie void is $\tau_{Mie} = 3.12$ ns (red line), clearly indicating accelerated decay dynamics. To exclude the influence of the local excitation field intensity on the apparent emission lifetime, we varied the excitation power and measured the decay dynamics of emitters within the Mie voids. In the case of a single-photon emitter, apparent lifetime depends on the excitation power, whereas in the limit of a large ensemble of emitters, their averaged lifetime is independent of the excitation power. Adjustment of the excitation power only changes the local field intensity while keeping the LDOS and emitter–resonator coupling unchanged, thus leaving the quantum-yield enhancement factor invariant. As shown in Figure 4G, under moderate power excitation (i.e., without causing rapid photobleaching), varying the excitation power does not affect the decay rate of the fluorophores, while the photoluminescence intensity exhibits a strictly linear dependence on excitation power. This allows the excitation enhancement term to be experimentally isolated from the quantum-yield enhancement contribution, providing direct validation of the local-field-driven excitation enhancement mechanism described by Eq. (1).

Meanwhile, the observed reduction in emission lifetime arises solely from resonant emitter–resonator coupling within the Mie voids, consistent with Purcell-enhanced decay channels introduced by the modified LDOS.

Fourier-plane emission imaging further allows us to probe the radiation channels associated with fluorophores within individual Mie voids. Angularly resolved emission spectra (Fourier spectrograms) measured on a bare silicon surface and within a Mie void along two orthogonal directions are shown in Figure 4H (corresponding panchromatic Fourier images are provided in Supplementary Figure S14). Compared with the bare silicon reference, the angular emission distribution within the collection range of $-20°$ to $20°$ remains largely unchanged in the presence of the Mie-void resonator. This observation suggests that the Mie-void resonator does not induce appreciable directional redistribution of the emitted photons or preferential coupling into specific radiation channels. Consequently, the experimentally observed photoluminescence modulation primarily reflects a genuine change in the emission intensity itself rather than an artifact arising from altered optical collection efficiency, i.e., the Mie-void resonance does not introduce significant radiation-channel engineering effects, and thus additional corrections associated with photon collection efficiency are not required in our analytical framework.

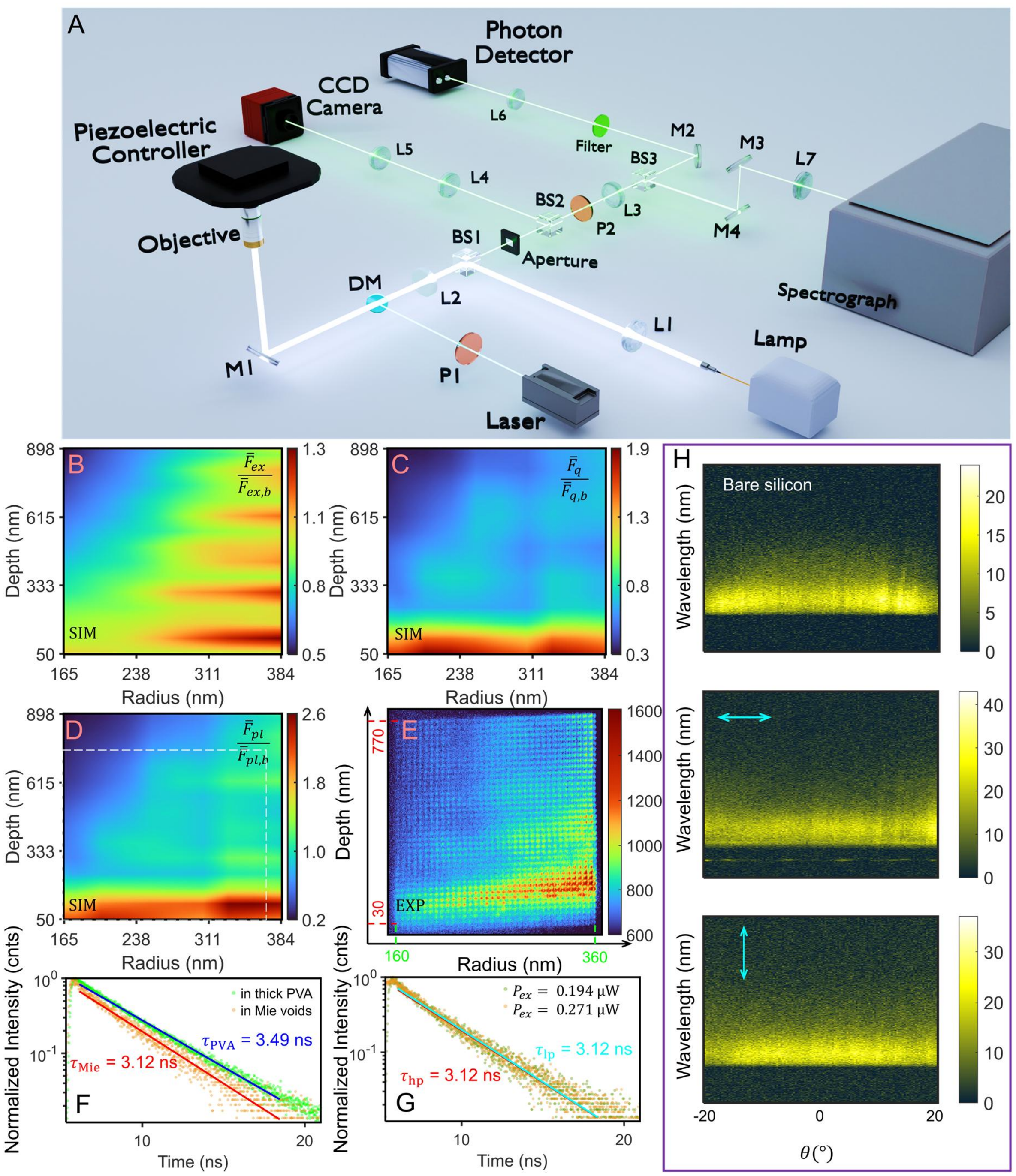

**Fig. 4. Theoretical prediction and experimental validation of photoluminescence tuning with Mie void. (A)** Schematic of the optical setup used for spectroscopic and TCSPC characterization. A 488 nm pulsed laser is synchronized with a time-resolved single-photon detector, and a digital piezoelectric positioning stage enables spatially resolved scanning of the sample. **(B–D)** Numerical simulations and calculations of the relative enhancement factors for cylindrical Mie voids with varying radius and depth, compared to a bare silicon substrate: (B) Relative excitation enhancement factor $\bar{F}_{ex}/\bar{F}_{ex,b}$; (C) Relative quantum yield enhancement factor $\bar{F}_q/\bar{F}_{q,b}$; (D) Combined relative photoluminescence tuning factor $\bar{F}_{pl}/\bar{F}_{pl,b}$, computed

using Eq. 4. The color maps reveal distinct geometric dependencies for excitation and emission processes, whose interplay governs the overall photoluminescence response. **(E)** Experimental photoluminescence maps obtained by scanning the Mie void gradient array. Fluorophores embedded within the PVA coating layer emit photons under pulsed excitation, and the resulting photon counts are spatially resolved. The region outlined by the white dashed box in (D) indicates the range of void geometries sampled in the experimental scan. A strong spatial correspondence between (D) and (E) confirms the predictive capability of the combined theoretical framework. **(F)** TCSPC measurements comparing the decay dynamics of emitters on the bare quartz surface (green data points with blue exponential fit, $\tau_{PVA} = 3.49$ ns) and within a single Mie void (orange data points with red exponential fit, $\tau_{Mie} = 3.12$ ns). The shortened emission lifetime reflects Purcell-assisted decay dynamics governed by Mie mode resonances. **(G)** Power-dependent TCSPC measurements of emitters within a Mie void, showing that the emission decay dynamics remain unchanged under varying excitation power, while the photoluminescence intensity scales linearly with excitation power. This confirms that excitation enhancement and emission dynamics are decoupled, with the latter governed solely by LDOS modification. **(H)** Fourier spectrograms of fluorophores embedded in a thin PVA layer on a bare silicon substrate (top) and within a Mie void for two orthogonal polarizations (middle and bottom). The angular distributions of emission from Mie voids exhibit no significant modifications within $\pm 20°$ as compared with the unpatterned surface. This confirms that the observed photoluminescence modulation arises from changes in excitation and emission processes rather than variations in optical collection efficiency.

Further, we fabricated a series of uniform Mie void arrays and experimentally characterized their optical response, including reflection spectra and void-tuned emission spectra of embedded fluorophores. As shown in Figure 5A, sample $S_0$ serves as a bare silicon reference without any nanostructures, and samples $S_1$ to $S_5$ are periodic arrays, each composed of 11 rows and 10 columns of identical Mie voids. While the layout remains constant, the geometrical parameters of the voids vary across the different samples, enabling controlled tuning of the photoluminescence properties. The reflection spectra for all five structured arrays are shown in Figure 5B. The observed spectral responses arise from the collective contributions of the uniform arrays, which in turn originate from the individual resonant characteristics of each void. Similarly, the photophysical properties observed from emitters within the arrays reflect the cumulative effect of resonant tuning at the single-void level. After spin-coating the arrays with a PVA layer uniformly embedded with fluorophores, we measured the emission spectra as shown in Figure 5C. Across all geometries, the Mie voids exhibit appreciable broadband resonance-enhanced fluorescence emission. Although Mie resonances yield lower enhancement factors compared to high-Q resonances such as Fano resonances or BICs[36,38,60], their broad resonant bandwidth enables robust spectral conformity, preserving the overall intrinsic emission profile of fluorophores and facilitating uniform enhancement across a wide wavelength range. Finally, the total emission intensity $I$ for each sample was obtained by integrating the corresponding fluorescence spectrum shown in Figure 5C. The relative enhancement was evaluated by normalizing to the bare substrate reference, defined as $I/I_b$, where $I_b$ represents the integrated emission intensity from fluorophores on the unstructured silicon surface. The experimental intensity enhancement ratios $I/I_b$ were plotted against the theoretical predictions $\bar{F}_{pl}/\bar{F}_{pl,b}$ , as presented in Figure 5D. While the experimental data points (fitted by a blue dashed proportional line) from the uniform Mie void arrays follow the overall trend predicted by the theoretical model, they consistently fall below the ideal unity-slope line (red dashed line), by ~5.4% on average. This modest deviation arises primarily from the presence of non-void regions within each periodic array, which contribute untuned baseline emission, as well as from experimental measurement uncertainties. Moreover, the depth of the Mie voids not only governs the resonance-mediated tuning of photoluminescence but also affects out-of-focus fluorescence collection, introducing further deviations between experimental observations and modeled results.

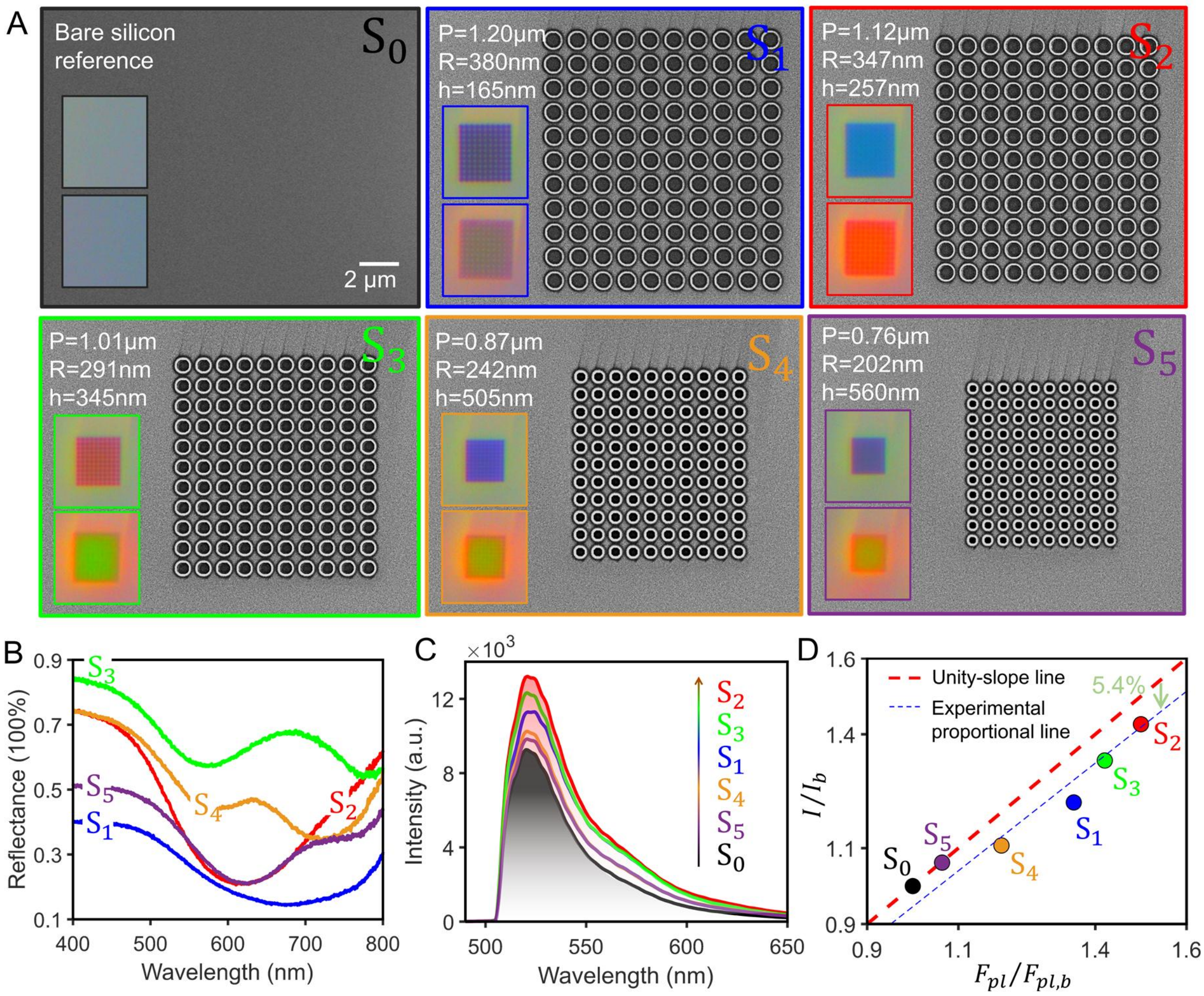

**Fig. 5. Photoluminescence measurement and characterization of uniform Mie void arrays.** **(A)** Scanning electron micrograph (right) and optical microscope images (top left: without PVA coating; bottom left: with PVA coating) of fabricated uniform Mie void arrays. All SEM images share the same scale bar. Sample $S_0$ serves as the bare silicon reference, while $S_1 - S_5$ correspond to periodic arrays of cylindrical Mie voids, each comprising an 11 × 10 grid with identical void dimensions. The arrays differ in void radius ($R$) and depth ($h$), while $P$ denotes the array period. **(B)** Measured reflectance spectra of the uncoated uniform Mie void arrays. Color labels correspond to samples $S_1 - S_5$. **(C)** Fluorescence emission spectra of the same arrays after spin-coating a PVA layer uniformly doped with fluorophores. Each Mie void geometry exhibits enhanced emission relative to the bare substrate, and the resonance-enhanced fluorescence preserves the intrinsic spectral profile of the fluorophores across the emission bandwidth. **(D)** Regression between theoretical and experimental photoluminescence enhancement. Experimental values are obtained by integrating the emission spectra in (C) and normalizing to $S_0$. The red dashed line indicates the ideal unity-slope correlation, while the blue dashed line represents the linear fit to the experimental data.

**High-Resolution Photoluminescence Encoding Pattern**

Having established the relationship between Mie-void geometry, spectral response, and photoluminescence modulation, we next exploit these design principles to realize a multimodal nanophotonic structure that encodes two high-resolution patterns within a single physical architecture while supporting three distinct display modalities, as illustrated in Figure 6. In this design, the EPFL logo is displayed under bright-field optical microscopy, while the SJTU logo remains hidden and emerges only under dark-field imaging or photoluminescence maps acquired via TCSPC scans. The spatial resolution of this pattern approaches the optical diffraction limit[61], with each pixel corresponding to a single Mie void with a submicron footprint. The pixel-level encoding scheme and corresponding Mie-void geometries are illustrated in Fig. 6A. Each void is assigned a two-bit optical code, yielding four distinct response states that can be selectively accessed through different imaging modalities. The first (least significant) bit determines the brightness state in the bright-field image — '1' for bright and '0' for dark — while the second (most significant) bit governs the brightness in both the dark-field image and the photoluminescence map. For instance, a pixel with the binary state '10' corresponds to a cylindrical Mie void designed with a diameter of 770 nm and a depth of 418 nm. This configuration appears as a dark pixel in the bright-field image but as a bright pixel under both dark-field and photoluminescence channels. The dimensions for the four states '00', '01', '10' and '11' are indicated in Figure 6A. Following this binary encoding strategy, the high-resolution nanophotonic pattern was fabricated via FIB milling, with tilted-view and top-view SEM images shown in Figures 6A and 6B, respectively.

We next characterized the fabricated patterns using optical microscopy under white-light illumination to assess their encoded optical responses. For reference, the optical responses of the four elementary Mie-void states — including reflected visible intensity, scattered visible intensity, and the photoluminescence tuning factor — were simulated, and the corresponding pixelated encoding patterns (Figures 6C, Figures 6D and Figures 6E) were compared with the experimental observations. Although subtle chromatic variations appear among the dark pixels in the experimental bright-field image of the uncoated pattern (Supplementary Figure S15A), grayscale conversion visually homogenizes the reflectance intensity of the dark-state pixels, thereby rendering a clearly defined EPFL logo in the bright-field channel (Figure 6F), in close agreement with the simulated result (Figure 6C). In contrast, the dark-field image of the uncoated sample reveals bright pixels with strong contrast against dark pixels, clearly forming the SJTU logo (Figure 6G), consistent with the simulation (Figure 6D). For comparison, Supplementary Section S10 presents additional ghost images with visual artifacts that arise when channels not explicitly engineered for independent image output are inadvertently activated. Moreover, photoluminescence mapping of the coated sample produces a high-contrast clear image, in which the SJTU logo emerges as bright pixels against a dark background (Figure 6H), in agreement with the simulation (Figure 6E), thereby demonstrating selective and pixelated tuning of photoluminescence by the Mie voids. This demonstration of ultracompact multimodal encoded patterns highlights the potential of Mie voids for spatially precise and spectrally selective control over optical reflectance, scattering, and photoluminescence. By integrating this resonator-enabled photoluminescence modulation mechanism with nanophotonic patterning, our approach achieves near–physical-limit resolution and enables multi-channel displays with independently programmable responses across bright-field, dark-field, and photoluminescence modalities. This capability establishes a new paradigm for high-density, high-fidelity optical information encoding and multiplexed encrypted display, paving the way for advanced applications in light-emitting metasurfaces, secure nanoscale tagging, and multiplexed photonic displays.

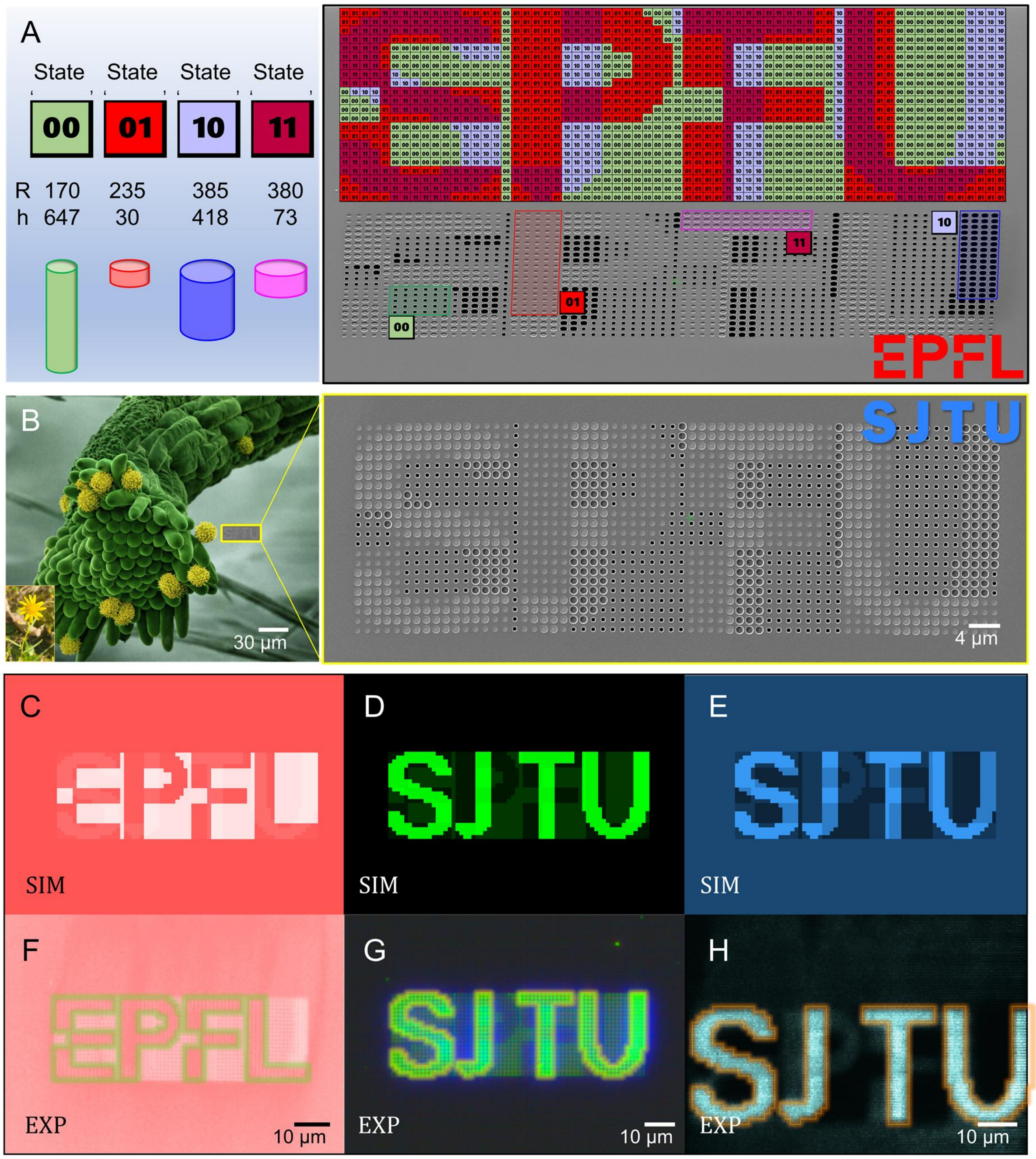

**Fig. 6. High-resolution photoluminescence encoding pattern displaying EPFL and SJTU logos. (A)** Binary encoding scheme and geometric design of individual pixelated Mie voids. Each state of void is assigned a two-bit code ($00 - 01 - 10 - 11$), with the corresponding radius ($R$) and depth ($h$) indicated below. The right panel shows the complete layout of the encoded pattern along with a tilted-view SEM image of the FIB-fabricated sample. **(B)** Top-view SEM image of the encoded-pattern sample. For an intuitive sense of scale, the entire pattern is juxtaposed with a dandelion stamen in the SEM, revealing an overall footprint comparable to a single pollen grain. The right panels of (A) and (B) are shown at the same scale. **(C–E)** Simulated pixelated encoding patterns for (C) bright-field, (D) dark-field, and (E)

photoluminescence channels. **(F)** Experimental bright-field image of the uncoated sample, in which the EPFL logo is clearly resolved. **(G)** Experimental dark-field image of the uncoated sample, directly rendering the SJTU logo. **(H)** Experimental photoluminescence map of the coated sample, showing a high-contrast SJTU logo defined by spatially selective emission enhancement at encoded Mie-void pixels.

## Discussion

Precise engineering of photoluminescence underpins the development of light-emitting devices, super-resolution imaging, optical information encoding and encrypted display technologies. However, achieving simultaneous and flexible control over local excitation enhancement and radiative quantum-yield modulation within an individual nanostructure unit remains challenging for conventional dielectric resonators, whose optical modes are predominantly confined within high-index media. In this work, we demonstrate air-defined silicon Mie voids as a flexible platform that provides emitters direct access to the regions of field confinement and Purcell enhancement, enabling tunable photoluminescence through the synergistic enhancement of localized excitation fields and resonance-mediated radiative decay. This behavior is quantitatively captured by an analytical framework that quantitatively distinguishes excitation and quantum-yield enhancement factors. Through full-wave simulations and experiments on FIB-fabricated Mie void structures, we observe Purcell-assisted lifetime shortening, geometry-dependent spontaneous emission orchestrating, and a near-linear correlation between the model-predicted tuning factor and measured fluorescence intensities. Building upon this experimentally validated model, we realize a multimodal nanophotonic pattern with near–diffraction-limit pixel resolution, in which each subwavelength Mie void encodes the EPFL logo in bright-field images and the SJTU logo in dark-field and TCSPC-based photoluminescence maps — achieving spatially independent, multichannel control without inter-pixel crosstalk. Compared with solid Mie particles, the complementary void geometry allows to achieve Purcell enhancement into low-index region, suppresses absorptive losses, thereby enabling spatially programmable and spectrally selective control at the single-resonator level. Beyond establishing a general methodology for the design of emitter–resonator architectures, these findings open new routes toward on-chip platforms for optical multiplexing encrypted display technologies, compact nanoscale tagging, and advanced light–matter interaction functionalities.

## Materials and Methods

### Fabrication

To realize the arrays of cylindrical Mie voids with independently varied depths and diameters on a silicon substrate (with a nominal resistivity of $\sim 1-10\ \Omega \cdot \mathrm{cm}$), we employed focused ion beam (FIB) milling to directly fabricate each nanovoid individually. Prior to nanostructuring, the silicon chips were subjected to Piranha cleaning to remove organic contaminants and particulates generated during wafer dicing. The Piranha cleaning protocol involved two identical and successive baths of concentrated sulfuric acid ($H_2SO_4$, 96%) heated at 100 °C and activated by hydrogen peroxide ($H_2O_2$, 30%). All nanostructuring was performed using a Zeiss CrossBeam 540 dual-beam system, which combines gallium ion beam and electron beam columns for high-resolution nanofabrication and imaging. This tool enables FIB patterning with a nominal resolution of $< 3$ nm. The void depth was precisely controlled by adjusting the ion exposure dose. A base dose of $1\ \mathrm{nC/\mu m^2}$ was defined, and a calibrated dose–depth relationship was experimentally established, revealing a near-linear correlation. The exposure dose for Mie voids was varied between 0.11 and 2.52 times the base dose, resulting in void depths ranging from 30 nm to 770 nm. All fabrication parameters for the nanovoid arrays—including spatial coordinates, diameters, and dose-adjusted depths—were generated via custom MATLAB R2024b scripts that output the complete milling instruction table. FIB milling was carried out in line-by-line scanning mode, with circular geometries exposed in a concentric outward pattern, using a 30 kV acceleration voltage (EHT) and a beam current of 100 pA.

### Characterization and Imaging

The microscale reflectance spectra of the periodic Mie void arrays were characterized using a custom-built inverted microscope system (Olympus IX73), equipped with a $20\times$ objective lens (numerical aperture = 0.30), and a spectrograph (Andor SR-303i-A) integrated with a Newton 971 EMCCD camera. The EMCCD camera provides real-time spatial imaging of the sample surface, allowing precise alignment and definition of the regions of interest on the sample surface for spectral acquisition. Broadband illumination was provided by a high-power supercontinuum white light fiber laser (SuperK FIANIUM, NKT Photonics), which was focused onto the nanostructured region via the objective. The reflected signal was collected through the same objective and directed into the spectrograph for wavelength dispersion and spectral recording. Reflectance spectra were obtained by normalizing the measured reflection from the structured sample to that of the unstructured bare silicon substrate. Bright-field and dark-field optical microscope images of the gradient array sample and encoded nanophotonic pattern were acquired using a Nikon Optiphot 200 inspection microscope equipped with a $20\times$, 0.40 NA objective lens. Illumination was provided by a halogen white light source, and the reflected light passed through a beam splitter before being captured by a CCD camera (Digital Sight DS-2Mv, Nikon). All optical images were processed solely by applying white balance correction taken from the bare silicon substrate, and no additional digital enhancement or filtering was applied.

The structural morphology of the Mie void arrays was characterized using field emission scanning electron microscopes (FE-SEM), including the Zeiss Crossbeam 550 (Carl Zeiss Jena GmbH) and Zeiss CrossBeam 540 systems, operated at an accelerating voltage (EHT) of 3.00 kV, beam current of 100 pA, and a working distance of 4.5 mm. High-resolution surface topography and void geometry were further characterized using atomic force microscopy (AFM; Bruker FastScan, equipped with a 5 MP digital camera), which provided scanning probe imaging with a nominal spatial resolution on the order of ångströms (Å), enabling precise quantification of individual void diameters and depths. Quantitative measurements of the void diameters and depths were performed and extracted by NanoScope Analysis software (version 1.50).

**Photoluminescence and lifetime measurement**

The acquisition of photoluminescence spectra and the measurement of emitter lifetimes were carried out using a custom-built experimental setup, as illustrated in Figure 4A. The lifetime measurements employed the time-correlated single-photon counting (TCSPC) method. An objective lens with 50 × magnification (NA = 0.9) was used. A slit was placed at the image plane of the tube lens (L2) to restrict the collected light to the region of interest. A beam splitter (BS2) diverted part of the signal to a Thorlabs CCD camera, with two additional lenses (L4 and L5) positioned in front of the camera to provide image magnification. White-light illumination was used to visualize the sample on the camera and was switched off during spectrum and lifetime measurements.

FITC (fluorescein isothiocyanate dissolved in dimethyl sulfoxide, 30 μmol/L embedded in 10% PVA) emitters were excited by a high-repetition-rate laser (LDH-D-C-488, peak wavelength: 488 nm), which was reflected by a dichroic mirror (DM) with a 500 nm cut-off wavelength. The resulting fluorescence emission, collected by the objective and peaking at ∼517 nm, passed through the dichroic mirror since its wavelength exceeded the cut-off. To suppress laser sidebands, a polarizer (P2) was placed orthogonal to the laser polarization, which was defined by polarizer P1. The light was then focused by the lens pair L3 and L7 onto the input slit of an Andor spectrograph, enabling precise emission spectrum measurements.

Fluorophore lifetime measurements were performed using a TCSPC system (PicoHarp300, PicoQuant GmbH), controlled via SymPhoTime64 software. The 488 nm pulsed laser was synchronized with a time-resolved single-photon detector. A digital piezoelectric stage enabled emission mapping by recording photon counts at each point within a maximum scan range and depth of ∼80 μm. Any point within the scanned region could be subsequently selected for lifetime acquisition. The lens pair L3 and L6 focused the collected light onto a single-photon avalanche diode (SPAD). Since the detector's active area was extremely small (∼50 μm), lens L6 was chosen to have a short focal length of 30 mm to achieve a tightly focused spot. A 520 nm bandpass filter (10 nm bandwidth) was used to isolate the fluorescence spectrum near the emission maximum. The measured decay curves were fitted with a single-exponential decay model to extract the emitter lifetimes.

**Numerical Simulation**

Geometric parameter sweeps of the Mie voids, including calculations of the near-field electromagnetic distributions, mode volumes, quality factors, far-field radiation energies, and reflection spectra, were conducted using the finite-difference time-domain (FDTD) method implemented in commercial electromagnetic simulation software (FDTD Solutions, Ansys Lumerical 2023 R1, Canada). The spatial integration of the near field, calculation of the Purcell factor, far-field radiation factor, and dielectric absorption factor were performed using custom

MATLAB scripts, based on the numerical near-field data and analytically defined expressions. The refractive indices and extinction coefficients of silicon substrate and the 10% PVA coating layer used in the FDTD models were experimentally determined by spectroscopic ellipsometry (J.A. Woollam RC2) and fitted with empirical dispersion models.

## Data availability

All data involved in this work are included in this article and the corresponding supplementary materials.

## Acknowledgments

The authors thank Prof. Harald Giessen (University of Stuttgart), Prof. Andrey Evlyukhin (Leibniz University Hannover), Dr. Stavros Athanasiou, and Dr. Parmenion Mavrikakis for fruitful discussions.

Funding from the Swiss National Science Foundation under grants 200021_212758 is gratefully acknowledged. We also acknowledge support from the National Natural Science Foundation of China under grants 62405177 and the Outstanding Doctoral Graduates Development Scholarship of Shanghai Jiao Tong University.

## Author contributions

Conceptualization: YF, OJFM;
Methodology: YF, IL, SB, MC, OJFM;
Investigation: YF, NZ, OJFM;
Visualization: YF, NZ;
Supervision: SB, SZ, OJFM;
Writing—original draft: YF, IL, OJFM;
Writing—review & editing: YF, IL, SB, TL, SZ, MC, OJFM.

## Competing interests

Authors declare that they have no competing interests.